\documentstyle [12pt]{article}
\def\ref{\par\noindent\hangindent 20pt}
\def\double{\baselineskip=24pt}

\begin{document}

\double

\begin{center}
{\bf DETECTING BIMODALITY IN ASTRONOMICAL DATASETS}
\end{center}

\begin{center}
K{\small EITH} M. A{\small SHMAN} and C{\small HRISTINA} M. B{\small IRD}\\
Department of Physics and Astronomy, University of Kansas, Lawrence,
KS 66045-2151\\
Electronic-mail: ashman@kusmos.phsx.ukans.edu, tbird@kula.phsx.ukans.edu\\
\end{center}
\begin{center}
and
\end{center}
\begin{center}
S{\small TEPHEN} E. Z{\small EPF}$^1$\\
Department of Astronomy, University of California, Berkeley, CA 94720\\
Electronic-mail: zepf@astron.berkeley.edu\\
\end{center}

\begin{center}
{\it To be published in the December 1994 issue of the Astronomical Journal}
\end{center}

\begin{center}
{\bf ABSTRACT}
\end{center}

We discuss statistical techniques for detecting and quantifying bimodality
in astronomical datasets. We concentrate on the KMM algorithm, which estimates
the statistical significance of bimodality in such datasets and
objectively partitions data into sub-populations. By simulating bimodal
distributions with a range of properties we investigate the sensitivity
of KMM to datasets with varying characteristics. Our results
facilitate the planning of optimal observing strategies for systems where
bimodality is suspected.
Mixture-modeling algorithms similar to the KMM
algorithm have been used in previous studies to partition the stellar
population of the Milky Way into subsystems. We illustrate the broad
applicability of KMM by analysing published data on
globular cluster metallicity distributions, velocity distributions of
galaxies in clusters, and burst durations of gamma-ray
sources. FORTRAN code for the KMM algorithm and directions for its
use are
available from the authors upon request.

\vfill

\noindent
$^1$ Hubble Fellow

\newpage

\begin{center}
{\bf 1. INTRODUCTION}
\end{center}

Many astronomical systems exhibit some form of clustering in their
constituents. This concept is most familiar in systems where the
clustering is spatial, such as in clusters of galaxies, but clustering
may be present in any variable. For instance, the metallicity bimodality
in the Milky Way globular cluster system (e.g.\ Zinn 1985) can be viewed as
clustering in metallicity.

Mixture modeling is a technique that can be used to detect clustering in
datasets and assess its statistical significance. The fundamental idea in such
an analysis is to fit models with different numbers of groups to a given
dataset and establish which model provides the best description of the
observations.
A thorough overview of the mixture-modeling approach and clustering analysis
is provided by McLachlan \& Basford (1988). The technique has been described in
an astronomical context by Nemec \& Nemec (1991, 1993) who concentrated on
the question of the number of distinct stellar populations in the Milky Way.

In this paper we concentrate on the use of mixture models to detect
bimodality in univariate datasets. Most of our discussion also applies
to the broader question of the detection of multimodality, but
an extension to multivariate datasets is beyond the scope of the present
paper.
In particular, we will describe a mixture-modeling algorithm known
as the KMM algorithm (Adams, McLachlan \& Basford 1993)
which can be used to detect the presence of two or more components
in an observational dataset. We present results on the ability of KMM
to correctly estimate the means and variances of two-Gaussian mixtures
and on the sensitivity of the likelihood ratio test to detect such
mixtures.

In addition to parameter estimation, the current version of the KMM code
includes a calculation of the {\it likelihood ratio test statistic}
(LRTS). Evaluation of the LRTS allows KMM to be used as a hypothesis
test. We study the specific case in which the variable of interest for
any single population has a Gaussian parent population. This is because
in many astrophysical problems, the variables of interest are
well-approximated by Gaussians. Examples are given in Section 4. [Agha
\& Ibrahim (1984) present a maximum likelihood method of parameter
estimation which allows a variety of functional forms of the
parent populations to be tested.] Our null hypothesis is that a unimodal
Gaussian parent population is a good description of the observational
data. Use of the LRTS allows the evaluation of the improvement in
goodness-of-fit for a two-component model relative to the unimodal; that is,
to test the null hypothesis.

Strictly speaking, the hypothesis test defined above is not identical to
testing for the presence of bimodality, at least in the case of finite
datasets. There are datasets that are better fit by two-groups
than a single Gaussian, for which a histogram of the data will not
show two clear peaks. This can arise either because of the finite nature of
the dataset, or because the parent population is comprised of two groups
which are not widely separated in location.
Since the presence of two distinct peaks in
such a histogram is often dependent on the {\it binning} of the data,
we regard our approach as more physically informative. From a statistical
point of view, the subjective nature of binning is unattractive and
it is preferable to
search for quantifiable clustering in the data rather
than attempting to locate two peaks in a histogram.

A related issue is the physical relevance that one places on bimodality.
A distribution with two distinct peaks can sometimes
be approximated by a quartic function,
so one can always claim that such a bimodal distribution consists of a single
population of objects with a probability density described by a quartic.
{}From such a viewpoint,
a mixture-model is nothing more than a mathematical tool for describing
a particular distribution. Our assertion is that in many astronomical
systems, key observable variables for individual populations
of objects have distributions that are consistent with a Gaussian.
Thus by demonstrating a significant improvement of a two-group
fit to a one-group fit we claim to ``detect'' the presence of more than
one group of objects in a given dataset.

The validity of this approach has been discussed by Nemec \& Nemec (1991)
in the context of the stellar populations within the Milky Way. They
recall a comment made by Eddington (1916) who noted that studying finite
mixtures of Gaussian distributions to model stellar populations provides
a crucial starting point for understanding the system. Having first
analyzed the system in this manner, it is then possible to add refinements
and study complications in the simple model. This is the spirit in which
we present the combination of mixture-modeling and the
LRTS that constitute the KMM algorithm.

A specific example may help to clarify this point. In the case of the
Milky Way, the metallicity of the globular cluster system has been
shown to be
bimodal (e.g.\ Zinn 1985). Armandroff \& Zinn (1988) fit two Gaussians
to the overall metallicity distribution, as did Nemec \& Nemec (1991)
using a mixture-modeling approach (see also Ashman \& Bird 1993).
As noted above, fitting the metallicity distribution in this
way can be viewed as nothing more than a mathematical convenience. However,
studies of the kinematics and spatial distribution of the Milky Way
globular cluster system (Zinn 1985; Armandroff \& Zinn 1988; Armandroff 1989)
have suggested that the two populations of globulars, as defined by the
two-group fit to the metallicity distribution, are indeed two distinct
populations. Thus verification of the presence of two populations of
Milky Way globulars, as indicated by their metallicity distribution, is
obtained through the study of additional variables.
This result provides much of the basis for the separation of Milky Way
globulars into disk and halo subsystems, as suggested by the earlier work
of Kinman (1959).

In many astronomical systems, only one
variable is easily accessible. In such cases, the detection of bimodality
in that variable provides evidence for the presence of two distinct
populations, but similar clustering in another variable is required to
make the identification more secure. Alternatively, if there are strong
theoretical reasons for believing that a variable should have a Gaussian
distribution [for example, the distribution of line-of-sight velocities
in a dynamically-relaxed cluster of galaxies (Bird 1993; see also Section 4.2
below)], then the presence of
bimodality provides powerful evidence that two distinct populations
are present.

The plan of this paper is as follows. In Section 2 we describe the KMM
algorithm and its application to univariate datasets. We also discuss other
statistical tests that can be used to explore a dataset for the presence
of bimodality. In Section 3 we present a study of the sensitivity of
 KMM when applied to
simulated bimodal datasets. KMM is applied to observational data in
Section 4 to illustrate the practical implementation of the algorithm
and its broad applicability to a range of astronomical systems.
We briefly discuss the application of the KMM
algorithm to multimodal models in Section 5.
Conclusions are presented in Section 6.

\begin{center}
{\bf 2. THE STATISTICAL TESTS}
\end{center}

\begin{center}
{\it 2.1 The KMM Algorithm and the Likelihood Ratio Test}
\end{center}

The KMM algorithm is an implementation of the EM (Expectation
Maximization) algorithm of
Dempster, Laird \&
Rubin (1977), described in detail by McLachlan \& Basford (1988).
Earlier work in this area is described by Orchard \& Woodbury (1972) and
includes the studies of Hasselblad (1966, 1969), Wolfe (1967, 1970) and
Day (1969).
KMM fits a user-specified number of Gaussian distributions to a
dataset, calculates the maximum likelihood estimates of their means and
variances and assesses the improvement of that fit over a single Gaussian.
The KMM algorithm assumes that each of the $N$
observed datapoints {\bf x$_j$} ($j = 1,2, ... ,N$) is
independently drawn from a parent population which is a mixture of Gaussian
probability densities.  The mixture density may be represented as
$$
f({\bf x};\phi )=\sum_{i=1}^g~\pi _i f_i({\bf x};\theta ).
\eqno(2.1)
$$
Here $\pi _i$ is the fraction of the parent population drawn from the $i$-th
component (also called the mixing fraction), $g$ is the total number of groups
or components being fit, $\theta$ is the vector representing all the unknown
model parameters for the $g$ components, and
$$
\phi = ({\bf \pi ',\theta '})'
\eqno(2.2)
$$
is the vector transpose of the unknown model parameters and mixing fractions.
 User input into the program (in addition to the measurements being
partitioned) is a first guess at
the means and covariances of the Gaussian distributions to be fit, as well
as an estimate of the mixing proportions.  The user
must also decide whether the routine should find groups with the same
covariances (homoscedastic fitting) or allow the groups to have arbitrary,
dissimilar covariances (heteroscedastic fitting).

An alternative option to providing the initial mixture of Gaussians is for
the user to provide group membership for the individual datapoints. In this
case, one can take advantage of techniques which estimate first splits of
the dataset (e.g.\ Kaufman \& Rousseeuw 1990). Based on this initial
 partition of the data, KMM calculates means and covariances of the
corresponding Gaussian components and proceeds as described below.
These two input options are essentially equivalent and we will only
refer to the former in what follows.

In its first pass through the data, KMM finds the best fit single Gaussian for
the given dataset (specifically, it finds the mean, $\mu$, and variance,
$\sigma ^2$, of the data).  The unimodal log(likelihood) is defined as
$$
L_C(1)=\sum _{j=1}^N~{\rm ln}~f(x_j;\theta )
\eqno(2.3)
$$
That is, it is the sum of the natural logarithms of the univariate
Gaussian evaluated at $x_j$.

Next, KMM enters its iterative step, using the value of $\phi$ defined in the
previous iteration (or specified by the user for the evaluation in the first
pass).  For each
datapoint it calculates the {\it a posteriori} probability that the object
belongs to each of the $g$ groups.  These probabilities are an estimate of
the confidence that the algorithm places in its assignment of any
particular object to any individual group.
The datapoint is assigned to the group for
which its probability of membership is the highest.  After the assignments are
made, the new vector $\phi$ is determined by finding the $\pi _i$, $\mu _i$ and
$\sigma ^2_i$ for each group.  The log(likelihood) of the $g$-group fit is
$$
L_C(\phi )=\sum _{i=1}^g~\sum _{j=1}^N z_{ij}~({\rm ln} \pi _i~+~{\rm ln}
f_i(x_j;\theta))
\eqno(2.4)
$$
Here $z_{ij}$ is an indicator variable.  In the ideal case, where the ``true''
external partition is known, $z_{ij}~=~1$ if object $j$ belongs to group $i$
and 0 otherwise.
KMM replaces this definition by the estimated posterior probability
 that object $j$ belongs to group
$i$.  This procedure is repeated
until the object assignments are stable and $L_C(\phi )$
has converged.
The algorithm converges to a local maximum
in the $\phi$-parameter space, provided that the sequence of likelihood values
is bounded above.  This is always true for mixture models with equal covariance
matrices for all components. In the heteroscedastic case, the sequence of
likelihood values may not be bounded above.

The final value of $L_C(\phi )$ is used to
determine whether or not the dataset is consistent with the null hypothesis.
The {\it likelihood ratio test statistic}  (hereafter LRTS)
is
$$
\lambda = L_C(\phi ) / L_C(1)
\eqno(2.5)
$$
and is an estimate of the improvement in going from a 1-mode to a $g$-mode fit.
The significance of the LRTS may be estimated by comparing --2ln$\lambda$ to
a $\chi ^2$ distribution with a number of degrees of freedom equal to
twice the difference between
the number of parameters of the two models under
comparison, not including the mixing proportions.
 However, this provides only an approximation of the statistical
significance. For the homoscedastic, univariate case, this approximation has
been shown to be adequate (McLachlan \& Basford 1988 and references
therein). However, for more complicated situations, the
only way to reliably assess the statistical significance
is a bootstrap estimation
(McLachlan 1987;
McLachlan \& Basford 1988; Nemec \& Nemec 1991;
Bird 1993; see also Section 5 below).
While bootstrapping individual moderate-sized datasets is trivial with current
workstations, bootstrapping thousands of simulated datasets to test KMM
sensitivity is too time-consuming. This is one reason why we restrict
attention in the present paper to univariate, homoscedastic datasets since
we can exploit the $\chi ^2$ distribution to obtain reasonably reliable
estimates of significance for large numbers of
simulated datasets.

The current version of the KMM code (Adams, McLachlan \& Basford
 1993) incorporates
an estimate of the statistical significance of the $g$-group fit over the
unimodal case using the analytical technique described above. The resulting
significance is expressed in terms of a $P$-value.
The $P$-value is the probability that the likelihood ratio test statistic
would be at least as large as the observed value if the null hypothesis
were true. In our restricted
case, the $P$-value represents the probability of determining the observed
value of the LRTS from a sample distribution drawn from a unimodal Gaussian.
The smaller the $P$-value, the less likely is the observed LRTS; that is,
small $P$-values indicate that the null hypothesis is not a good description
of the data. Conventionally, $P<0.05$ indicates that the null hypothesis
is strongly inconsistent with the data, whereas $0.05<P<0.1$ indicates
marginal inconsistency.

In addition to the $P$-value, the KMM algorithm returns values for the
estimated means of the groups, $\mu _i$, the estimated
mixing fraction, $\pi _i$,
and the estimated
common covariance, $\sigma ^2$. (In the heteroscedastic case, the
covariance for each group is calculated.) The algorithm also assigns
individual datapoints to the groups for which their posterior probability
is highest, and estimates a confidence value for each assignment.
Thus KMM provides not only a
powerful hypothesis test, it also allows objective partitioning of
data into individual groups for a given $g$-group fit.

\begin{center}
{\it 2.2 Double-Root Residuals}
\end{center}

Double-Root Residuals (hereafter DRRs) provide an effective way of comparing
any univariate dataset with a model distribution. Their use in an
astronomical context has been described by Gebhardt \& Beers (1991).
Calculating residuals to compare a model with observed data and displaying the
results as a difference plot (or hanging histogram) is a common technique.
However, the drawback of this method is that residuals in
heavily-populated bins are often much larger than those in bins with few
datapoints, simply because $\sqrt N$ fluctuations
are larger. Consequently, the distribution of residuals is often a poor
diagnostic of the deviation of the observed data from the model distribution.

The DRR plot represents a simple refinement of the more familiar difference
plot that overcomes the shortcomings inherent in ordinary residuals. By
taking square-roots of both the number of observed datapoints per bin and
the corresponding model prediction, the $\sqrt N$ problem described above is
effectively removed. Gebhardt \& Beers (1991) note that an improvement
over a comparison of simple square roots is provided by the expression:
$$
{\rm DRR} = \sqrt{2 + 4N_{data}} - \sqrt{1 + 4N_{model}}~~(N_{data} \ge 1)
\eqno(2.5a)
$$
$$
{\rm DRR} = 1 - \sqrt{1 + 4N_{model}}~~(N_{data} = 0)
\eqno(2.5b)
$$
(cf.\ Velleman \& Hoaglin 1981). Here $N_{data}$ is the number of datapoints in
a given bin and $N_{model}$ is the number of points in a given bin as
predicted by the model under investigation.

The DRR plot illustrates
clearly exactly {\it where} the observed data and model differ significantly
from one another. In cases where the model provides an adequate fit to the
data, DRRs also have the useful property of having numerical
values roughly equivalent to normal deviates. Thus DRRs can be used to test
the fit of a model to a given dataset. If the DRR has
a local value of 2 or greater, this indicates a discrepancy between the model
and the data at the 2$\sigma$ (95\%) level.
We illustrate these properties explicitly in Sections 3 and 4.

Using DRRs to test the validity of a unimodal versus bimodal distribution
is a relatively straightforward procedure. If, as in the KMM case, one
is testing the hypothesis that the data is well-fit by a single Gaussian,
one calculates the best-fit Gaussian to the data, bins that Gaussian,
and calculates the DRRs using equation (2.5).
It is a simple matter to modify the
model distribution against which the data is to be tested. One way this
can be exploited in the detection of bimodality
is to compare the data with a uniform distribution. The rationale here is that
``boxy'' distributions (which have fewer datapoints in the tails of the
distribution relative to a Gaussian)
are generally better fit by a mixture of Gaussians
than a single Gaussian. Thus there is a concern when fitting Gaussian
mixtures that a two-group model will be preferred over a one-group model
{\it not} because the parent population is comprised of two components,
but because it is intrinsically light-tailed. As we illustrate in Section 4,
calculating the DRRs for a uniform model can help determine whether the
parent distribution is genuinely bimodal or simply boxy.

The primary drawback of the DRR
technique is that it
relies on binning of the data. The likelihood ratio test exploited by
the KMM algorithm is free of this
weakness. Nevertheless, it is also apparent that DRRs are a useful addition to
the arsenal of statistical weapons that can be used to detect or reject
the presence of bimodality.

\begin{center}
{\it 2.3 Shape Estimators and ROSTAT}
\end{center}

The ROSTAT statistics package (Beers et al.\ 1990; Bird \& Beers 1993) is
designed to determine robust estimates of the location, scale, and shape of
univariate datasets. ROSTAT contains an array of statistical
tests, some of which can be exploited to detect bimodality and mixtures of
distributions in a dataset.

We retain our earlier assumption that
individual components in a dataset can be approximated by Gaussian
distributions. A mixture of two (or more) Gaussians
in which the individual means of the two components differ produces a
distribution that is light-tailed.
If, in addition, the
individual components of a distribution contain significantly different
numbers of datapoints, the result is usually an asymmetric distribution
in which more points lie on one side of the mean than the other. Examples
of datasets with these properties are discussed in Sections 3 and 4.

Traditionally, the shape of a distribution is described by its
kurtosis and skewness (cf.\ Bird \& Beers 1993 and references therein).
The skewness, which is the third moment about the mean of a distribution,
measures the degree of asymmetry of the distribution and is given by:
$$
a_3 = {{ \int (x - \mu)^3 f(x) dx } \over {\sigma ^3}}
\eqno(2.6)
$$
where $f(x)$ is the probability density function
under study, $\mu$ is its mean, and
$\sigma$ is its standard deviation.
The kurtosis is the fourth moment about the
mean of a distribution:
$$
a_4 = {{\int (x - \mu)^4 f(x) dx } \over {\sigma ^4}} - 3
\eqno(2.7)
$$
With these definitions, the skewness and kurtosis are both equal to zero
for a Gaussian distribution.

Bird \& Beers (1993) have compared the skewness and
kurtosis with robust shape estimators known as the asymmetry and tail
indices. They find that, while the asymmetry and tail indices are
more powerful for comparing the shapes of univariate distributions, the
skewness and kurtosis are preferable if one is primarily concerned with
testing the hypothesis that a given distribution is consistent with
Gaussian. In the context of the present discussion, we suggest that
skewness and kurtosis provide a useful diagnostic in the search for
possible bimodality. The ROSTAT package returns the probability that
the dataset under study is consistent with a Gaussian distribution,
as determined by these quantities. This complements the
other methods described earlier in this Section, although we emphasize
that of these three techniques only the likelihood ratio test combined with
mixture modeling directly addresses the issue
of multiple components in a dataset.

It is important to stress that the techniques described above represent
only a fraction of the tests that can be applied when investigating
the presence of bimodality. A fuller discussion is provided by
McLachlan \& Basford (1988), Hartigan \& Hartigan (1985)
 and Nemec \& Nemec (1991). However, we have
found that the KMM algorithm, with support from DRRs and the normality
tests of the ROSTAT package, is a powerful
tool for studying the question in practice.

\begin{center}
{\bf 3. TESTING THE SENSITIVITY OF KMM}
\end{center}

We have carried out a series of experiments to study the sensitivity of the
KMM algorithm to bimodality.
Specifically, we have generated realizations of bimodal distributions
consisting of two equal-variance Gaussians
in which the mixing proportions are equal. For computational convenience,
we fixed the mean of group 1 at
${\mu}_1=0.0$ and varied ${\mu}_2$ between 1.75 and 3.25 in steps
of 0.25. The common
variance of the groups was set at ${\sigma}^2=1.0$ and the total number
of data points, $N$, varied from 50 to 500.

We define a dimensionless separation of the means:
$$
{\Delta}{\mu} = \frac{ ({\mu}_2 - {\mu}_1)}{ \sigma }
\eqno(3.1)
$$
With our assumed values, it is apparent that
${\Delta}{\mu}={\mu}_2$. Thus varying ${\mu}_2$ as described above corresponds
to varying ${\Delta}{\mu}$. For each
pair of ($N$,${\Delta}{\mu}$) we generated 100 simulated datasets. The KMM
algorithm was then applied to these distributions.

Before discussing our results, it is worth stressing the analogy between
our numerical experiments and the application of KMM to observational
data. First, there are analytic results concerning the
detectability of bimodality for idealized double-Gaussian distributions
 with equal
mixing proportions.
One of the simplest and most
useful is that such a distribution only shows two peaks if:
$$
{\Delta}{\mu} \ge 2.0
\eqno(3.2)
$$
(cf.\ Everitt \& Hand 1981).
In other words, in the case of infinite $N$, a distribution will have two
peaks if equation (3.2) is satisfied. With the exception of the
cases with $\Delta \mu = 1.75$, all of the simulated distributions
are drawn from such idealized parent populations.
However, we expect KMM to return different
significance levels of bimodality on distributions even when they are
drawn from a parent population with the same ($N$,$\Delta \mu$), simply
because of the finite number of datapoints.
The important point to note here is that {\it observation} of a finite
number of objects drawn from some parent population is directly analogous to
simulating distributions. Thus studying the performance of KMM on
our simulations provides a direct measure of the detectability of
bimodality using KMM on observed astronomical datasets, provided of course that
the observed datasets have similar statistical properties to the simulated
ones.

As described in Section 2, the KMM algorithm requires user-specified starting
values for the estimated means and common variance of the groups to
be fit. When running KMM on the simulated datasets, we provided it
with the values used to produce the simulated datasets.
We have found that for two-group homescedastic fitting, KMM is
insensitive to these input values and will converge to the ``correct''
2-group fit even when its starting values are far removed from the
true means and variance (see also Everitt 1981, Bird 1993).
(The only exceptions are starting values
in which either the means are very close together or the variance is
very large.) Thus while using these values as input speeds
up the convergence of KMM, it does not affect the results we obtain.
When applying KMM to real datasets it does not hurt to try a range of
starting values, but in our experience the two-group homoscedastic
fit is forgiving of poor initial choices.

The results of running KMM on these simulated distributions are summarized
in Tables 1 and 2 and Figures 1 to 3. For our present study,
the most important quantity returned
by the KMM algorithm is the $P$-value of the
hypothesis test that a two-group fit is an improvement over a one-group
fit (see Section 2.1 above).
In Table 1 we present the median $P$-value, $\langle P \rangle$,
based on 100 simulations for each parameter pair ($N$,${\Delta}{\mu}$).
The median value of $P$ provides a more robust estimate of the
``characteristic'' $P$-value than a straight mean, primarily because the
distribution of $P$-values for any given parameter pair tends to be
highly skewed (see Section 3.1 below). The information in Table 1 is
displayed graphically in Figure 1.

Our experiments revealed that for $N<50$ the likelihood ratio
test used in the KMM algorithm does {\it not}
provide a reliable method for detecting bimodality. For $N=50$ it is
apparent that large separations in the means of the two components are required
for KMM to ``typically'' return a strongly significant ($P<0.05$) rejection
of the unimodal hypothesis. As $N$ gets larger, the likelihood of rejecting
the unimodal hypothesis increases. Note that for the parameter pair ($N~=~500$,
${\Delta}{\mu}~=~1.75$) we obtain
$\langle P \rangle~<~0.05$, despite the fact that this value of $\Delta \mu$
lies below the bimodality constraint in equation (3.2). This partly reflects
the fact that with finite $N$ one can be fortunate and draw a sample that
is bimodal even when the parent distribution does not satisfy (3.2).
However, this also reflects the fact
that KMM can accurately partition data into two groups and demonstrate
the improvement of fit over a unimodal distribution even when the parent
distribution is not bimodal. Thus while it is not entirely accurate
to describe such distributions as bimodal, one can test the likelihood
that the parent distribution is comprised of two distinct groups.
In astronomical applications, the presence of two components is of more
physical significance than the presence of two distinct peaks (cf.\
our comments in Section 1
above).

In Table 2a and 2b we present the fraction of simulations for each
parameter pair in which
$P < 0.05$ and $P < 0.1$, respectively. Figures 2a and 2b display
this information graphically.
Since the distributions are generated from double-Gaussian
parent populations,
these results can be viewed as the efficiency with which KMM correctly
identifies the presence of two Gaussian components.
(Note that one reason this efficiency
is not 100\% is because in some of
these finite realizations the double-Gaussian
characteristics are washed out. Again, this is analogous to the process of
observation
when, with finite datasets, one can simply be unlucky.)

In Figure 3 we display the information contained in Tables 2a and 2b
in the ($N$,$\Delta \mu$)-plane. The four lines represent loci of
the fixed fractions of simulations (99\% and 95\%) for which
$P < 0.05$ and $P < 0.1$. The lines are simple linear interpolations  of
the results given in Tables 2a and 2b.

As one would intuitively expect, KMM becomes more efficient at detecting
bimodality as $N$ and $\Delta \mu$ increase (see also McLachlan 1987;
Mendell et al.\ 1992, 1993). The quantitative results
can be exploited when planning observational strategies. For instance,
if one is attempting to test a model in which bimodality is predicted to
exist in an observable quantity at some $\Delta \mu$, Table 2a and Figure 2a
give the number of datapoints required to detect such bimodality with a
given probability.

These figures and tables can also be used in the interpretation of
analyses of real data. This may be particularly important when the user
has been forced to throw out datapoints in an attempt to remove
a background or other contamination of the dataset.
Such clipping of
the data is often justified for physical or
statistical reasons and is sometimes necessary to obtain reliable
results with KMM.  For instance, a ``3$\sigma$--clipping'' routine was
proposed by Yahil \& Vidal (1977) to identify galaxies as cluster members
based on their redshifts.  While their iterative technique has been shown
to be unreliable (Beers et al.\ 1990), redshift filters are
regularly applied to galaxy cluster velocities to reduce foreground and
background contamination (cf. Malumuth et al.\ 1992, Bird 1994).

If one retains a constant background in
a dataset, the result is usually a distribution with artificially
extended tails.
In such datasets, KMM will attempt to fit these tails (as well as
potentially-significant deviations) when finding the
optimum one- and two-group fit, thereby producing spurious results.
In fact, under such conditions, the significance of the two-group
fit will usually
be reduced, since a single, broad unimodal typically provides
a better fit to
the long tails. The potential hazard in clipping univariate data
is that such truncation tends to make a dataset more boxy. This
generates the opposite effect to a constant background contamination, and
in some instances it is possible
that the likelihood of the two-group fit will be artificially
increased.

One way to guard against this problem is to use the results of the
sensitivity tests presented above. If one only obtains a
strong bimodal signal after clipping a dataset,
 a comparison of the value of $N$ and the derived
value of ${\Delta} {\mu}$ with Figures 2 and 3 may help determine whether
the bimodality signal is an artifact of the clipping procedure.
If the observed values lie in a region of the ($N$, $\Delta \mu$)-plane with
a low expectation of rejecting the unimodal hypothesis, it is likely
that clipping has artificially introduced the bimodal signal. We return to
this point in Section 4.

\begin{center}
{\it 3.1 Distribution of $P$-values, Means, and Variances}
\end{center}

To illustrate the typical dispersion in the parameters returned by KMM, we
now study one set of simulations in more detail. Specifically, we look
at the 100 simulated datasets with $N = 300$ and $\Delta \mu = 2.25$.
Inspection of other datasets reveals that the distribution of the
KMM output
parameters becomes more concentrated around the input values as one goes
to larger $N$ and $\Delta \mu$, as expected.

In Figure 4 we show that distribution of $P$-values returned by KMM for
the simulations with $N = 300$ and $\Delta \mu = 2.25$. While most
of the values strongly reject the unimodal hypothesis, there is a
tail of high $P$-values which do not reject the unimodal model.

Figures 5(a)--(e) show the distributions
 of the other parameters returned by KMM.
Recall that the input parameters for the simulations here have the mean
of the first group $\mu _1 = 0.0$, the mean of the second group
$\mu _2 = 2.25$, the common variance $\sigma ^2 = 1.0$ (so that
$(\mu _2 - \mu_1)/\sigma ^2 = 2.25$) and the ratio of mixing
proportions, $\pi _1/\pi _2 = 1.0$. Thus these figures can be viewed as a
summary of how well KMM is doing in returning the input values (although,
as noted in Section 2, much of the variation in the returned parameter
values reflect the finite nature of the simulated datasets).

Figures 5(a) and 5(b) show the distribution of $\mu _1$ and $\mu _2$,
respectively. More than half the returned values lie within 0.1 of the input
value.
In Figure 5(c) we show the distribution of
$\sigma ^2$ returned by KMM, which is clearly sharply peaked around the
input value of 1.0. Figure 5(d) shows the distribution of separation of the
means normalized to the value of $\sigma$ returned by KMM. In Figure 5(e)
we show $\pi _1/\pi _2$ returned by KMM.
These figures show that the parameter values calculated by
KMM are reliable estimates of the parameters describing the parent
populations.

\begin{center}
{\it 3.2 A Case Study}
\end{center}

We now look in detail at one randomly selected dataset drawn from the 100
simulations of the $N = 300$, $\Delta \mu = 2.25$ set discussed in the previous
subsection. The purpose is to compare the results returned by KMM with other
statistical tests and to illustrate the application of the ROSTAT package and
DRRs to such datasets. A histogram of this dataset is given in Figure 6, where
the best-fitting single Gaussian ($\mu = 1.21$, $\sigma = 1.5$) is also
shown.

One of the most striking aspects of Figure 6 is that the histogram does
not look particularly bimodal! This is somewhat surprising, given the large
number of points and respectable separations of the means of the two
groups. Despite the appearance of the histogram, the KMM algorithm separates
the distribution into two groups at a high degree of confidence. Specifically,
KMM returns $\mu _1 = -0.074$, $\mu _2 = 2.250$, $\sigma ^2 = 0.914$, and
$\pi _1/\pi _2 = 128/172$. The rejection of the unimodal hypothesis is
strongly significant with $P = 0.001$. Comparison with the results summarized
in
Section 3.1 above show that this is an unexceptional dataset from the $N =
300$,
$\Delta \mu = 2.25$ set of simulations. The ratio $\pi _1/\pi _2$ is on the low
side, but the other results are typical for these input parameters.

Perhaps the most notable
result is that KMM correctly identifies, at strong
significance, the presence of two groups in this dataset, even though the
histogram of the distribution does not have two obvious peaks. This
illustrates the superiority of an objective algorithm such as KMM over binning
the data and assessing the results by eye. However, it is worth noting
that Figure 6 reveals the boxiness which is characteristic of some two-group
mixtures.

The normality tests contained in the ROSTAT package return interesting
information on the dataset, which further
 supports the results of the KMM analysis.
Firstly, the skewness has a value of --0.104, indicating a deficiency
of datapoints with values less than the mean.
This level of skewness does {\it not} reject the single Gaussian hypothesis.
(This is encouraging since
although the parent population is not Gaussian it {\it is} symmetric.)
On the other hand, the kurtosis for this datset has a value of --0.665,
corresponding to a rejection of the single Gaussian hypothesis
at a confidence level of $P < 0.001$. Negative kurtosis is expected for
a bimodal
dataset and supports the rejection of the single
Gaussian model by KMM.

Finally, we apply the DRR technique to the dataset in Figure 6. In Figure
7(a) we plot the DRRs of the dataset when compared to the best-fitting single
Gaussian model. The residuals creep above the
2$\sigma$ level on each side of the single Gaussian mean,
in the manner expected if the distribution is bimodal. Thus the DRR plot
rejects the single Gaussian model at 95\% confidence.
In Figure 7(b) we plot DRRs for
a model with a uniform distribution. Here one has a certain flexibility in
deciding how far the uniform distribution extends. However, the presence
of highly significant DRRs with both negative and positive sign indicates
that a uniform distribution is inconsistent with this dataset.

In other datasets, particularly when $N$ is small, we have found that uniform
distributions are often acceptable fits to genuine bimodal distributions,
according to the DRR technique.
This is primarily because of the boxy nature of bimodal distributions when
binned. In Figure 7(b), the largest positive DRR is produced by the
heavily-populated bin at $x=2.0$. Creative binning can significantly
reduce this bin and hence the associated DRR, although for this particular
dataset we have not been able to find a sensible binning that reduces all
DRRs below 2$\sigma$. As noted earlier, this dependence on binning is an
unattractive aspect of the DRR method. We therefore advocate the use of
KMM as the primary diagnostic unless there are good reasons for believing
that the parent
distribution of the variable of interest for a single population is
non-Gaussian.

\begin{center}
{\it 3.3 Playing Devil's Advocate}
\end{center}

The results discussed above indicate that the likelihood ratio test,
using KMM partitions, can correctly identify bimodality
in univariate datasets. However, it is also of interest to establish the
frequency with which KMM erroneously ``detects'' bimodality in datasets
drawn from single Gaussian parent distributions.
To investigate this possibility we simulated unimodal Gaussian distributions
with $N$ varying from 50 to 500 as before. The mean of the Gaussian was set
at 0.0, with a dispersion of 1.0. For each value of $N$ we
generated 100 simulations and ran KMM on the datasets. We provided KMM
with the input values: $\sigma ^2 = 1.0$, $\mu _1 = -0.5$, $\mu_2 = 0.5$.
 The results of this
exercise are presented in Figure 8, which shows
the fraction of the simulations $f_{05}$ and
$f_{10}$ which returned $P$-values less than 0.05 and 0.10, respectively.

It is apparent from Figure 8 that KMM occasionally identifies bimodality
in samples drawn from a Gaussian parent population.  However, even in
low-$N$ datasets, the false-positive frequency is never greater than 10\%.
Such a level is expected on the basis of the definition of the $P$-value.
About 50\% of the false bimodality detections in these simulations can be
rejected using the kurtosis returned by ROSTAT. That is, for about
half of these
distributions the kurtosis indicates consistency with a single Gaussian.
Thus using KMM in conjunction with the kurtosis of the
dataset leads to $\sim$5\% false detection
of bimodality for $N=50$, with the fraction dropping at higher $N$.
Given the power of the KMM algorithm in correctly identifying bimodality,
we conclude that this low level of false detections is acceptable.

\begin{center}
{\bf 4. APPLICATION TO DATA}
\end{center}

\begin{center}
{\it 4.1 The Color Distribution of Globular Clusters around NGC 5128}
\end{center}

Broad-band colors of old globular clusters can be used to deduce
their metallicities (e.g., Harris 1991 and references therein).
The shape of such metallicity distributions contains valuable information
about the formation of globular cluster systems and the early
chemical and dynamical evolution of their host galaxies. Of interest
to the present discussion is the suggestion that a merger origin
for elliptical galaxies leads to a globular cluster metallicity
distribution which is bimodal (Ashman \& Zepf 1992). More generally,
two or more populations of globular clusters around a galaxy indicates
that the formation of these objects probably occurred in discrete
bursts rather than through some continuous process.

In the Milky Way, the halo and disk globular cluster systems
individually have metallicity distributions
 that are well-approximated
by Gaussians (cf.\ Armandroff and Zinn 1988; Nemec \& Nemec 1991).
As noted in Section 1, kinematic and spatial distributions of Milky
Way globulars confirm this separation of the system into two
subsystems. It is therefore reasonable to hypothesize that individual
populations of globular clusters have a distribution in [Fe/H] that is
Gaussian-distributed. In metallicity distributions derived from
broad-band photometry, errors in [Fe/H] (arising from internal photometric
errors) are also Gaussian. This is therefore an ideal situation to
use the KMM algorithm and its assumption that individual populations
have Gaussian-distributed variables (in this case, [Fe/H]).

The Washington $(C-T_1)$ index has good sensitivity to metallicity
variations and has been used to derive metallicity distributions
for a number of extragalactic globular cluster systems. As an example,
we consider the color distribution of globulars around NGC 5128
presented by Harris et al.\ (1992). [A summary of the results from
a KMM study of this system was given by Zepf \& Ashman (1993).]

In Figure 9(a) we show the distribution of $(C-T_1)$ colors for the
globular clusters around NGC 5128. These are the colors
presented by Harris et al.\ (1992) without any correction for
reddening. It is apparent that two of the datapoints are somewhat
removed from the main distribution. Both these extreme points are flagged
as uncertain by Harris et al.\ (1992), the blue object having a close
neighbor and the red one being near the NGC 5128 dust lane. Moreover,
both colors are beyond the range for which the metallicity-color relationship
is calibrated. Since the foundation of the Gaussian hypothesis here is that
{\it metallicity} is Gaussian-distributed, these two extreme points
could compromise the analysis. We therefore exclude them from the
following discussion. The curve shown in Figure 9(a) is the best-fit
single Gaussian to the remaining 60 datapoints.

A KMM study yields the following results. The $P$-value for the dataset
is 0.093, representing a marginally significant rejection of the
unimodal hypothesis. The means of the two groups
are 1.53 and 1.95, respectively, with a common variance of
$\sigma ^2 = 0.024$. KMM assigns 33 objects to the blue group and
27 to the red.
We noted in Section 3 that comparing KMM output with the results from
simulated datasets provides a useful check of whether the KMM output
is reasonable. This has additional merit in cases such as the present one
where a couple of points have been excluded from the KMM analysis.
The KMM output parameters imply $\Delta \mu = 2.7$ for this $N = 60$
dataset. From Figures 1 and 2 it is apparent that a $P$-value of
0.093 is typical for a dataset with these values of $\Delta \mu$ and
$N$. This provides added support that KMM is accurately partitioning this
dataset and that the exclusion of the two extreme datapoints is
reasonable.

Zepf \& Ashman (1993) have suggested that the
blue colors of some clusters in NGC 5128 may be indicative of youth
rather than low metallicity.
In particular,
 Zepf \& Ashman (1993) noted that the metallicity-radius relation for
globulars in this galaxy is more consistent with other systems
if the blue globulars at small projected radii are
young rather than metal-poor. The two bluest clusters in the 60-point
dataset both have small projected radii.
If they are removed from the analysis, the bimodality signal in this
system is much stronger. A KMM analysis of this 58-point dataset returns
a $P$-value of 0.004, corresponding to a strong
rejection of the single-Gaussian hypothesis. The means and mixing proportions
are not significantly affected by this further clipping (the variance of
the two groups drops somewhat), providing some evidence that the partitions
are reasonable. However, this does illustrate the subjective element
that inevitably creeps in to such analyses, as well as the slightly
unstable performance of KMM in datasets with small $N$. On a more
optimistic note,
the KMM analysis has provided direction for future observations which
could definitively establish whether this globular cluster color
distribution is bimodal.

In Figure 9(b) and 9(c) we show the DRR plots of the best uniform
and Gaussian model, respectively. The Gaussian model is consistent at
the 2$\sigma$ level with the data. This result agrees with the KMM
analysis where the rejection of the single Gaussian was at less
than 95\% confidence. The uniform distribution formally is rejected
at greater than 2$\sigma$, although it is apparent from Figure 9(c)
that the discrepancy is caused by the tails of the dataset. Since it
is possible that datapoints could be scattered into outlying bins
through observational uncertainties, this is not a compelling
reason to reject the uniform distribution. However, the arguments
given above that individual globular cluster populations have
Gaussian metallicity distributions suggest to us that the bimodal
interpretation of this data is more reasonable.

The kurtosis of this dataset is --0.737, corresponding to a $P$-value of 0.072
and a marginally significant rejection of the Gaussian hypothesis. This
is encouragingly similar to the KMM result. Note that the kurtosis is
negative, as expected for a bimodal distribution. The skewness shows
no significant departure from Gaussian, indicating a symmetric dataset
(as expected for a bimodal distribution with a similar number of points
in each peak). Other examples of the application of the KMM algorithm
to globular cluster color distributions are given by Ashman \& Zepf
(1993), Ashman \& Bird (1993), Lee \& Geisler (1993), Ostrov et al.\
(1993), Zepf \& Ashman (1993) and Zepf, Ashman \& Geisler (1994).

\begin{center}
{\it 4.2 Galaxy Velocities in the Cluster A548}
\end{center}

The line-of-sight velocities of galaxies in a dynamically-relaxed
 cluster are expected
to be Gaussian-distributed
(Bird 1993 and references therein).
Deviations from a Gaussian distribution may indicate
that the cluster is unrelaxed, that substructure is present, or
the three-dimensional velocity distribution is anisotropic. A
double-Gaussian
distribution is suggestive of two gravitationally-distinct groups of
galaxies. Clearly, this
situation is another example where the KMM algorithm can provide useful
information about the physical state of the system, since the assumption
of individual populations having Gaussian-distributed velocities has a firm
foundation.  While the positions of galaxies also provide important
insight into the dynamical structure of these massive systems, for the
purposes of this paper we will discuss the application of KMM to velocity
distributions only.  KMM has been applied simultaneously to velocity and
position information for clusters of galaxies by Bird (1994) and Davis et al.\
(1994).

The KMM algorithm and other techniques have been applied to velocity
datasets of many clusters of galaxies (Yahil \& Vidal 1977; Beers et al.\ 1990;
Beers et al.\ 1992;
Bird \& Beers 1993; Zabludoff, Franx and Geller 1993).
To demonstrate the application
of KMM to a cluster velocity dataset, we consider the cluster A548, which
was observed by Dressler \& Shectman (1988) in their study of substructure
in rich clusters of galaxies.
In Figure 10(a) we show the velocity
distribution of 133 galaxies along with the best-fitting Gaussian
to the dataset. Velocities are given relative to the cluster
recessional velocity of 12394 km/s (Davis et al.\
1994).
The dataset is clearly somewhat more boxy than a
Gaussian, providing a hint that bimodality might be present.

The results of applying the KMM algorithm provide strong evidence
that the galaxy velocities have at least a bimodal distribution.
The $P$-value of 0.003 rejects the single Gaussian model.
The parameters calculated by KMM for the two-group
homoscedastic model include individual group means of --629 km/s and
$+$771 km/s and a common variance of (517 km/s)$^2$.
The two groups are
similarly populated with 71 points assigned to the low-velocity group
and 62 to the high-velocity group.

Figures 10(b) and 10(c) show the DRRs for the best-fitting Gaussian and
a uniform distribution, respectively. The Gaussian model has both
negative and positive DRRs in excess of 2, ruling out the model at the
2$\sigma$ level. In an attempt to fit a uniform distribution,
we excluded the tails of the dataset (which would give large negative DRRs).
Even with this helping-hand, the DRRs for the uniform distribution
shown in Figure 10(c) clearly rule out the uniform model.

The kurtosis of this dataset confirms its boxy appearance with a value
of --0.860 and a $P$-value less than 0.001, strongly rejecting
the Gaussian model. The skewness shows consistency with a symmetric
distribution,
again reflecting the roughly equal number of points in the two groups.

Although the dataset shown in Figure 10(a) does not look strongly bimodal,
the results of these tests indicate clearly that a two-group fit is
preferred over a single group fit. The larger number of points in this
dataset relative to the one considered in the previous subsection is
an important factor in producing
this higher level of confidence in the analysis.

The combination of velocities and positions for galaxies in A548,
presented along with {\it ROSAT} X-ray observations of this
system in Davis et al.\ (1994),
suggest that its physical structure is even more
complex than the KMM analysis indicates.  When the location of gas
density peaks are used to constrain the number of groups that KMM
fits, A548 is found to possess at least four physically-distinct
subunits, of which three are probably gravitationally-bound.  In
this case, the use of KMM provides a powerful addition to the
gas and galaxy diagnostics provided directly by the observations.

\begin{center}
{\it 4.3 The Duration of Gamma-Ray Bursts}
\end{center}

Recent work by Kouveliotou et al.\ (1993) has shown that the duration of
gamma-ray bursts (hereafter GRBs) has a bimodal distribution. These authors
further show that the burst durations are anticorrelated with spectral
hardness, adding support to the claim of two distinct populations. Since
the nature of GRBs remains controversial, we do not have strong theoretical
reasons for assuming any particular form for the distribution of
 burst durations.
However, the logarithmic distribution of $T_{90}$ has the appearance
of two roughly Gaussian distributions [see Figure 11(a) of this paper
 and Figure 1(a)
of Kouveliotou et al. (1993)]. Here, $T_{90}$ is the time in seconds
 during which
the cumulative BATSE (Burst and Transient Source Experiment on the
{\it Compton Gamma-Ray Observatory}) background-subtracted signal
increases from 5\% to 95\% of the total counts.
Like Kouveliotou et al.\ (1993) we use the data contained
in the First BATSE Catalog (Fishman et al.\ 1994). This contains 222
datapoints with reliable duration measurements.

We ran the KMM algorithm on log($T_{90}$) and obtained a rejection of a single
Gaussian with $P<0.001$. The means of the two groups occur at
log($T_{90}) = -0.346$ and 1.462 (or 0.45s and 28.9s, respectively).
KMM assigns 63 points to the short duration group and 159 to the long
duration group. This is consistent with the visual appearance of
Figure 11(a) that the two groups contain a significantly different number of
objects. The variance in log$(T_{90})$ is (0.242s)$^2$.

DRR plots are shown in Figures 11(b) and 11(c). Both the Gaussian and
uniform model fits are strongly rejected. The kurtosis of the dataset is
--0.692, corresponding to $P<0.001$ for the single Gaussian model.
The skewness also rejects a single Gaussian. Its value
of --0.606 reflects the smaller number of objects in the short duration
group.

\begin{center}
{\bf 5. DISCUSSION}
\end{center}

In the above Sections we have
demonstrated the ability of KMM to evaluate the presence of
two-group mixtures of Gaussians.
There are many astronomical applications where the two-group
homoscedastic model is theoretically and/or empirically plausible.
However, there are also situations where there is little empirical or
theoretical motivation for believing the homoscedastic assumption
is valid.
In such cases it may be useful to drop the equal variance requirement and
use KMM to look for heteroscedastic groups. In galaxy cluster applications,
for instance, heteroscedastic fits can be very helpful for isolating small
foreground and background groups [e.g., the cluster A85, Bird (1994)].

One disadvantage of heteroscedastic mixture models is the lack of a
reliable analytic
approximation to the significance level. As discussed in Section 2,
the $\chi ^2$ distribution is only a good approximation to the
$P$-value distribution for bimodal homoscedastic fits (McLachlan \&
Basford 1988). In other cases, the bootstrap technique must be employed.
Generally speaking, any bootstrap algorithm generates
non-independent random datasets under the null hypothesis of the statistic
being evaluated. It allows a determination of the significance of a
statistic (in the present case, the likelihood ratio test statistic) based
on the sampling fluctuations and distribution properties of the observed
dataset. [See Efron (1981), McLachlan \& Basford (1988);
Beers et al.\ (1990) and Bird (1993) for detailed discussions
of bootstrapping techniques.]

Extensive experimentation with stellar and galaxy cluster datasets
suggests that the estimated $P$-value provided by the KMM algorithm usually
{\it underestimates} the signifance of a given model (Bird 1993). That is,
in most cases the bootstrap $P$-value is even lower (i.e., {\it more}
significant) than the analytically estimated value.
Thus the analytic approximation is useful when one is exploring a dataset
by fitting different heteroscedastic models, in that it provides a guide to
preferred models. Such studies can avoid the time-consuming
process of bootstrapping $P$-values for a large number of different models.

Using KMM to detect multimodality ($g>2$)
 in homoscedastic datasets suffers from
similar limitations. The analytic $P$-value is often a useful guideline
to the significance of a given model, but bootstrapping is generally
required to obtain an accurate value. Comparing a $g$-group homoscedastic
fit to a unimodal Gaussian can be carried out in a fashion analogous to the
2-group versus unimodal fit described in earlier Sections.
The main uncertainty is the determination of the optimum number
of groups for a given dataset. This question has been the focus of extensive
statistical research in recent years [see, for instance, Roz\'al \&
Hartigan (1994) and references therein]. Unfortunately, no single powerful
technique has yet emerged for determining how many groups are present
in a dataset.

Nemec \& Nemec (1991) advocate an Occam's razor
approach in which the optimum number of groups is determined by the smallest
number of groups that is consistent with the data. These authors
discuss various techniques of how the goodness-of-fit of such
mixture models can be assessed and illustrate these techniques by
applying them to stellar populations within the Milky Way.

A similar method has been suggested by Bird (1994) who obtains analytic
$P$-values for $g$-group fits versus the unimodal case for a range of $g$.
If a number of models return strongly significant rejections of the
unimodal hypothesis, Occam's razor suggests that the fit with the smallest
number of groups be preferred. Other criteria can be added, such as
rejecting models where one or more of the groups contains a small fraction of
the total number of datapoints. (For example, this may be applicable if
small groups are deemed unlikely on physical grounds.)
It is, of course, still possible to use
the LRTS to directly compare the significance of two competing models
by bootstrapping $P$. In practice, observed
datasets frequently do not strongly point to a unique solution.
Ultimately, there is no substitute for
additional empirical information to determine a {\it physically}
preferable partition.

There are statistical techniques that address some of these issues
more directly, many of which are described by McLachlan \& Basford (1988).
Of particular interest is the procedure proposed by Hawkins (1981) which
tests simultaneously for homoscedasticity and normality of the components
of a mixture model. Since our primary aim in this paper is to present
the KMM algorithm, we refer the interested reader to McLachlan \& Basford
(1988), Hawkins (1981) and Fatti, Hawkins \& Raath (1982) for further
information on Hawkins' test and other techniques. More recently,
Roeder (1994) has introduced a graphical technique for identifying
the number of groups in a mixture of Gaussians.

The above comments and caveats indicate that KMM has many uses beyond the
bimodality hypothesis test discussed in earlier Sections. However, it
is also apparent that in the univariate, homoscedastic, bimodal case,
the hypothesis test is relatively straight-forward and easy to implement.
With more complicated models, more care must be used when applying KMM,
but even in such cases much can be gleaned by exploring the clustering
within a dataset using KMM.

\begin{center}
{\bf 6. CONCLUSIONS}
\end{center}

We have discussed a variety of tests for detecting and quantifying
bimodality in astronomical datasets. The most important of these is a
mixture-modeling technique implemented using the KMM algorithm.
Our simulations indicate that this algorithm provides a powerful means
to explore bimodality in univariate datasets. We have illustrated the
practical use of KMM by applying it to data from three astrophysical
systems.
The primary goal of
this paper is to introduce the astronomical community to
the KMM algorithm as applied to univariate datasets.
FORTRAN code for the
KMM algorithm is available electronically from the authors upon request,
along with further instructions on the use of the algorithm.

\bigskip

\bigskip

We are very grateful to Timothy Beers for introducing us to the KMM
algorithm and for a number of helpful conversations. We also owe a
considerable debt to Geoff McLachlan and Peter Adams
for educating us in the finer points of mixture-modeling and cluster
analysis. We
thank Barbara Anthony-Twarog, Mike Nowak and Bruce Twarog for reading
an earlier version of this paper and for their helpful contributions.
 K.M.A.
gratefully acknowledges the financial support of a Fullam/Dudley Award
and a Dunham Grant from the Fund for Astrophysical Research. C.M.B.
is supported by NSF Grant No.\ OSR-9255223 at the University of Kansas.
S.E.Z. acknowledges the support of
NASA through grant number
HF-1055.01-93A awarded by the Space Telescope Science Institute, which
is operated by the Association of Universities for Research in
Astronomy, Inc., for NASA under contract NAS5-26555.

\begin{center}
{\bf REFERENCES}
\end{center}

\ref{Adams, P.A., McLachlan, G., \& Basford, K. 1993, private communication}

\ref{Armandroff, T.E. 1989, AJ, 97, 375}

\ref{Armandroff, T.E., \& Zinn, R. 1988, AJ, 96, 92}

\ref{Ashman, K.M., \& Bird, C.M. 1993, AJ, 106, 2281}

\ref{Ashman, K.M., \& Zepf, S.E. 1992, ApJ, 384, 50}

\ref{Ashman, K.M., \& Zepf, S.E. 1993, in The Globular Cluster---Galaxy
Connection, eds Brodie, J., \& Smith, G., (Dordrecht: Kluwer),
p.776}

\ref{Beers, T.C., Flynn, K., \& Gebhardt, K.  1990, AJ, 100, 32}

\ref{Beers, T.C., Gebhardt, K., Huchra, J.P., Forman, W., \& Jones, C.
1992, ApJ, 400, 410}

\ref{Bird, C.M. 1993, PhD Thesis, University of Minnesota}

\ref{Bird, C.M. 1994, AJ, 107, 1637}

\ref{Bird, C.M., \& Beers, T.C. 1993, AJ, 105, 1596}

\ref{Davis, D.S., Bird, C.M., Mushotzky, R.F., \& Odewahn, S.C.
1994, ApJ, submitted}

\ref{Day, N.E. 1969, Biometrika, 56, 463}

\ref{Dempster, A.P., Laird, N.M., \& Rubin, D.B. 1977, J. R. Statist. Soc.
B 39, 1}

\ref{Dressler, A. \& Shectman, S. 1988, AJ, 95, 985}

\ref{Eddington, A.S. 1916, MNRAS, 77, 16}

\ref{Efron, B.  1981, {\it Biometrika}, 68, 589}

\ref{Everitt, B.S.  1981, Multivar.\ Behav.\ Res.\, 16, 171}

\ref{Everitt, B.S., \& Hand, D.J. 1981, in Finite Mixture Distributions
(New York: Chapman \& Hall)}

\ref{Fatti, L.P., Hawkins, D.M., \& Raath, E.L. 1982, in Topics in
Applied Multivariate Analysis, ed D.M. Hawkins (Cambridge University Press),
p.~1.}

\ref{Fishman, G.J., et al. 1994, ApJ Supp, in press}

\ref{Gebhardt, K., \& Beers, T.C. 1991, ApJ, 383, 72}

\ref{Harris, G.L.H., Geisler, D., Harris, H.C., \& Hesser, J.E. 1992,
AJ, 104, 613}

\ref{Harris, W.E. 1991, ARAA, 29, 543}

\ref{Hartigan, J.A., \& Hartigan, P.M. 1985, Ann. Statist. 13, 70}

\ref{Hasselblad, V. 1966, Technometrics, 8, 431}

\ref{Hasselblad, V. 1969, J.\ Amer.\ Statist.\ Assoc.\ 64, 1459}

\ref{Hawkins, D.M. 1981, Technometrics, 23, 105}

\ref{Kaufman, L., \& Rousseeuw, P.J.
1990, in Finding Groups in Data: An Introduction
to Cluster Analysis (New York: Wiley)}

\ref{Kinman, T.D. 1959, MNRAS, 119, 559}

\ref{Kouveliotou, C., Meegan, C.A., Fishman, G.J., Bhat, P.N., Briggs, M.S.,
Koshut, T.M., Paciesas, W.S., \& Pendleton, G.N. 1993, ApJLett, 413, L101}

\ref{Lee, M.G., \& Geisler, D. 1993, AJ, 106, 493}

\ref{Malumuth, E.M., Kriss, G.A., Dixon, W.V.D., Ferguson, H.C. \& Ritchie,
C.  1992, AJ, 104, 495}

\ref{McLachlan, G.J. 1987, Applied Statistics, 36, 318}

\ref{McLachlan, G.J., \& Basford, K.E. 1988, in Mixture Models: Inference and
Applications to Clustering (New York: Marcel Dekker)}

\ref{Mendell, N.R., Finch, S.J., \& Thode, H.C. 1993, Biometrics, 49, 907}

\ref{Mendell, N.R., Thode, H.C., \& Finch, S.J. 1992, Biometrics,
47, 1143 (erratum: Biometrics 48, 661)}

\ref{Nemec, J.M., \& Nemec, A.F.L. 1991, PASP, 103, 95}

\ref{Nemec, J.M., \& Nemec, A.F.L. 1993, AJ, 105, 1455}

\ref{Orchard, T., \& Woodbury, M.A. 1972, in Proceedings of the 6th Berkeley
Symposium (Vol.\ 1) (Berkeley: University of California Press), p.~697}

\ref{Ostrov, P., Geisler, D., \& Forte, J.C. 1993, AJ, 105, 1762}

\ref{Roeder, K. 1994, J.\ Amer. Stat.\ Assoc.\, 89, 487}

\ref{Roz\'al, G., \& Hartigan, J.A. 1994, Journal of Classification, in press}

\ref{Velleman, P.F., \& Hoaglin, D.C. 1981, in Applications, Basics, and
Computing of Exploratory Data Analysis (Boston: Duxbury Press)

\ref{Wolfe, J.H. 1967, Research Memo, SRM 68-2 (San Diego: U.S.\ Naval
Personnel Reasearch Activity)}

\ref{Wolfe, J.H. 1970, Multivar.\ Behav.\ Res.\ 5, 329}

\ref{Yahil, A., \& Vidal, N.V. 1977, ApJ, 214, 347}

\ref{Zabludoff, A.I., Franx, M., \& Geller, M.J. 1993, ApJ, 419, 47}

\ref{Zepf, S.E., \& Ashman, K.M. 1993, MNRAS, 264, 611}

\ref{Zepf, S.E., Ashman, K.M., \& Geisler, D. 1994, in preparation}

\ref{Zinn, R. 1985. ApJ, 293, 424}

\newpage

\begin{center}
{\bf FIGURE CAPTIONS}
\end{center}

\noindent
{\bf Figure 1.} Median $P$-value versus $\Delta \mu$
for a range of $N$, as determined by the KMM algorithm when
testing for a 2-group fit over a 1-group fit.

\medskip

\noindent
{\bf Figure 2.} The fraction of simulations
 for which KMM returns:
(a) $P~<~0.05$; (b) $P~<~0.1$ plotted against
$\Delta \mu$. The different lines represent different
values of $N$. Symbols are the same as in Figure 1.

\medskip

\noindent
{\bf Figure 3.} The
frequency of strongly significant ($P~<~0.05$) and
marginally significant ($P~<~0.1$) rejection of the unimodal hypothesis
in favor of a 2-group fit. Loci are shown for 95\% and 99\% rejection
frequencies.

\medskip

\noindent
{\bf Figure 4.} The distribution of $P$-values as returned by KMM for
100 simulations with $N = 300$ and $\Delta \mu = 2.25$. The range of
$P$-values shown is: (a) $0.000 < P < 0.01$; (b) $0.01 < P < 0.25$.

\medskip

\noindent
{\bf Figure 5.} The distribution of parameters returned by KMM for 100
simulations with $N = 300$ and $\Delta \mu = 2.25$: (a) mean of first
Gaussian, $\mu _1$; (b) mean of second Gaussian, $\mu _2$; (c) common
variance, $\sigma ^2$; (d) $(\mu_2 - \mu_1)/\sigma$; (e) ratio of
mixing fractions, $\pi _1/\pi _2$.

\medskip

\noindent
{\bf Figure 6.} A histogram showing one realization of the $N = 300$,
$\Delta \mu = 2.25$ simulations. This is the dataset studied in detail
in Section 3.2. The best-fitting single Gaussian for this dataset
is superimposed.

\medskip

\noindent
{\bf Figure 7.} DRR plots for the dataset shown in Figure 6 when compared to
(a) a Gaussian; (b) a uniform distribution.

\medskip

\noindent
{\bf Figure 8.} Fraction of simulations of unimodal distributions in which
KMM returns $P < 0.1$ ($f_{10}$) and $P < 0.05$ ($f_{05}$). Only partitions
with at least 20\% of the datapoints in each group are included.

\medskip

\noindent
{\bf Figure 9.} (a) The $(C-T_1)$ colors of globular clusters around
NGC 5128 from the data of Harris et al.\ (1992).
The superimposed curve is the best-fitting Gaussian to the 60-point
dataset in which the
two outliers have been removed. (b) The DRRs of the
the best-fitting Gaussian to the 60-point dataset. (c) The DRRs of a uniform
model distribution.

\medskip

\noindent
{\bf Figure 10.} (a) The velocity distribution of galaxies in the cluster
A548 from Dressler \& Shectman (1988). The best-fitting Gaussian is
superimposed. (b) The DRRs of the best-fitting Gaussian model.
(c) The DRRs of a uniform model distribution with the two high-velocity
datapoints removed.

\medskip

\noindent
{\bf Figure 11.} The distribution of log$(T_{90})$ of BATSE gamma-ray
burst durations from the data of Fishman et al.\ (1994). The
best-fitting Gaussian is also shown. (b) The DRRs of the best-fitting
Gaussian model. (c) The DRRs of a uniform model distribution.

\end{document}